\documentclass[12pt, a4paper]{article}

\usepackage{fullpage}
\usepackage{type1cm}

\usepackage{authblk}
\usepackage{makeidx}     
\usepackage{graphicx}    
              
\usepackage{multicol}    
\usepackage[bottom]{footmisc} 

\usepackage{newtxtext}    %
\usepackage{newtxmath}    
\usepackage{color}
\usepackage{hyperref}

\makeindex       

\newcommand{\quotes}[1]{``#1''}
\newcommand{\mm}[1]{{#1}}

\graphicspath{{./figures/}{./images/}}
\usepackage{authblk}

\title{Quantifying exaptation in scientific evolution}

\author[1, 2]{\small M\'arcia~R. Ferreira}
\author[1]{\small Niklas~Reisz}
\author[1]{\small William~Schueller}
\author[1]{\small Vito D.P.~Servedio}
\author[1, 3, 4, 5]{\small Stefan~Thurner}
\author[1, 6, 7]{\small Vittorio~Loreto}

\affil[1]{\footnotesize Complexity Science Hub Vienna, Josefst\"adter Str. 39, 1080 Vienna, Austria}
\affil[2]{\footnotesize TU Wien (Vienna University of Technology), 1040 Vienna, Austria}
\affil[3]{\footnotesize Section for Science of Complex Systems, Center for Medical Statistics, Informatics and Intelligent Systems (CeMSIIS), Medical University of Vienna, A-1090 Vienna, Austria}
\affil[4]{\footnotesize International Institute for Applied Systems Analysis (IIASA), A-2361 Laxenburg, Austria}
\affil[5]{\footnotesize Santa Fe Institute, Santa Fe, NM 85701}
\affil[6]{\footnotesize Sony Computer Science Lab, Paris, 6, Rue Amyot, 75005 Paris, France}
\affil[7]{\footnotesize Sapienza University of Rome, P.le Aldo Moro 2, 00185, Rome, Italy}

\begin{document}

\maketitle
\abstract{
\noindent
Rediscovering a new function for something can be just as important as the discovery itself. In 1982, Stephen Jay Gould and Elisabeth Vrba named this phenomenon \emph{Exaptation} to describe a radical shift in the function of a specific trait during biological evolution. While exaptation is thought to be a fundamental mechanism for generating adaptive innovations, diversity, and sophisticated features, relatively little effort has been made to quantify exaptation outside the topic of biological evolution. We think that this concept provides a useful framework for characterising the emergence of innovations in science. This article explores the notion that exaptation arises from the usage of scientific ideas in domains other than the area that they were originally applied to. In particular, we adopt a normalised entropy and an inverse participation ratio as observables that reveal and quantify the concept of exaptation. We identify distinctive patterns of exaptation and expose specific examples of papers that display those patterns. Our approach represents a first step towards the quantification of exaptation phenomena in the context of scientific evolution.\footnote{This article is a preprint version of results of a study published in an upcoming Springer-Nature: The Frontiers Collection book entitled: \quotes{Understanding innovation through
exaptation} edited by Stefano Zapperi and Caterina La Porta.}}

\hfill\break\noindent\textbf{Keywords:} Exaptation, Evolution, Innovation, Scientometrics, Science of Novelty

\section{Introduction}

The discovery of a new function or application for a given object or concept can be just as important as the discovery of something for the first time. This phenomenon is known as \emph{exaptation}, and the related verb is \emph{to co-opt}. It characterises the process of acquiring new functions for which a trait, which originally evolved to solve one problem, is co-opted to solve a new problem \cite{gould1982exaptation}. The definition is similar to the concept of \emph{preadaptation}~\cite{bock1959preadaptation}. However, since this term may suggest teleology, Vrba and Gould urged scholars to replace that term by exaptation. The idea of exaptation was also proposed to distinguish the concept from \emph{adaptation}~\cite{darwin2004origin}. While exaptations are traits that have applications that deviate from their original purpose~\cite{gould1982exaptation,kauffman2000investigations}, adaptations have been shaped by natural selection for their current use~\cite{bock1959preadaptation,darwin2004origin}. One canonical example of exaptation in biology is the evolution of feathers.  It is often argued that feathers were not originally developed for flight, but emerged in the reptilian ancestors of today's birds for thermal regulation~\cite{gould1982exaptation}.  \mm{Other examples include the ability of a metabolic reaction network to survive on different food sources which can allow adoption of alternative substrates \cite{barve2013latent}.}
Exaptation events seem to be important to give birth to adaptive innovations, diversity and, more generally, to complex traits~\cite{barve2013latent}.

Gould and Vrba propose exaptations have adaptive and non-adaptive origins: \emph{preaptation} and \emph{nonaptation}. Preaptation refers to characteristics or traits that are adapted and `selected' for one evolutionary purpose (adaptations). These are later co-opted to serve another purpose leading to an increase in the fitness of the co-opted trait. Thus, preaptations are adaptations that have undergone a significant change in function~\cite{gould1982exaptation,lloyd2017exaptation}. Another source of exaptation is nonaptation. Nonaptation refers to traits that emerge through a process of trial and error that generates lots of `leftover' features (e.g., DNA, molecules, cells)~\cite{gould1982exaptation}. Nonaptation concerns the effective use of co-opted leftover traits to serve a particular function, but whose origin cannot be ascribed to the process of `natural selection' \cite{gould1982exaptation}. Thus, nonaptations are byproducts of the evolution of some other trait ~\cite{darwin2004origin} that do not add to the fitness of the co-opted trait \cite{lloyd2017exaptation,gould1982exaptation}. In summary, adaptations, preaptations, and nonaptations are essential processes that drive the evolution of living matter, cells, humans, organisms, and biological ecosystems.  They allow us to understand the adaptive and non-adaptive origins of biological novelty.

Recently, the notion of exaptation has been applied to the study of technological change \cite{bonifati2010more,dew2004economic}. One set of studies have focused on the development of technological speciation narratives~\cite{dew2004economic,levinthal1998slow,andriani2013exaptation} and niche construction theory~\cite{dew2016exaptation}. Other studies focused on the adoption of technology~\cite{rogers2010diffusion}, its commercial application~\cite{schumpeter1939business}, and its economic impact~\cite{dew2016exaptation}, but not on the origin of those inventions \cite{fleming2004science}. Small-scale studies of technological exaptation abound in the management and innovation literature (e.g., \cite{tan2015alexander, dew2016exaptation, cattani2005preadaptation,rosenman1988serendipity}).
\mm{These studies point to the role of chance such as serendipitous discovery of a new function \cite{andriani2013exaptation}. Serendipity (accidental circumstances leading to fortunate findings)  and exaptation (a shift in the function of something) are intricately related by the fact that accidental discoveries that contribute to the redeployment of a component in a different context lead to a shift in the function of that component.}
A related stream of research has attempted to model the dynamics of invention mathematically by analysing specific knowledge spaces~\cite{tria2014dynamics,loreto2016dynamics,thurner2010schumpeterian,klimek2010evolutionary,klimek2012empirical,hidalgo2009building,tacchella2012new,servedio2018new,kauffman1993origins, gabora2013quantum}. The aim of these studies is not to explain why some entities produce more innovations than others (productivity), or what influences the ability of these entities to produce them, but how knowledge evolves in a mechanistic view. 
These studies have provided evidence for the existence of innovation bursts in national economies~\cite{thurner2010schumpeterian,klimek2012empirical}, the rediscovery of publications leading to the emergence of new scientific fields~\cite{thurner2019role,van2004sleeping}, and the novel combination of components as a source of everyday novelties~\cite{tria2014dynamics}. 
Few attempts focused on explicitly quantifying exaptation, one notable exception being~\cite{andriani2017measuring}, where it is estimated that about 40\% of pharmaceutical drugs have started as something else. 

Similarly, many scientific discoveries find applications that are not foreseen from the outset. 
\mm{The scientific context is particularly relevant for the study of exaptation since it encompasses a variety of human activities where knowledge is frequently rediscovered and re-used.}
In scientific evolution processes, different disciplines may come together, \quotes{to tell one coherent interlocking story}~\cite{watson2017convergence}, or a field may subdivide into new disciplines. Both may form the basis on which concepts can further recombine. The recent use of statistical physics to examine people's behaviour in crowds, traffic, or stock markets is an example of co-opting theories and techniques to the social sciences. 
Research on laser technologies~\cite{bonifati2010more}, pharmaceuticals~\cite{andriani2017measuring}, and fibre optics~\cite{cattani2005preadaptation} keep expanding their scope of application in very diverse fields. 
This kind of repurposing may enhance (though not necessarily) the fitness of entities~\cite{gould1982exaptation}. 
Exaptation in the context of science thus refers to a diversification logic, where publications build on an existing knowledge base and succeed in entering other fields by creating new scientific niches or sub-fields. 
We thus interpret exaptation in science as how research insights from publications in one field are co-opted (i.e., cited) by publications from different scientific domains. 

To arrive at a formal understanding of exaptation in science and technology we start by briefly reviewing related perspectives on the origins of innovations.%
\footnote{It is not our aim to provide an in-depth discussion and definition of the concepts of novelty, innovation, or invention, which can be found elsewhere~\cite{erwin2004innovation,arthur2007structure}; these terms will be used interchangeably throughout the paper.}
We then attempt to detect the fingerprints of exaptation by using publication data indexed in the APS (American Physical Society) database, which contains over 450,000 articles in the field of physics. We use the direct citation network to construct clusters of publications that represent sub-fields of physics. Citations have been considered as a proxy for academic relevance, with citation-based indicators offering approximate information about the scientific impact of publications~\cite{sugimoto2018measuring}.
\mm{We focus on seminal publications that initially appear in a given domain and later receive acknowledgements and new functions from other domains. The idea of function is critical here because it reflects knowledge-use of specific research outcomes in different scientific contexts.} 
By investigating the \quotes{citation paths} of individual publications across different domains, we quantify both their overall importance, as well as their impact on the structure of the field of physics as a whole.

The paper is structured as follows. 
In Section 2, we introduce the concept of exaptation as a theoretical concept and clarify the scientific context in which we will use it in this paper. 
In Section 3, we describe the data and the empirical approach. 
In Section 4, we present results. 
Finally, in Section 5, we conclude \mm{the paper and summarise our results and proposal in this area of research.} 

\subsubsection*{\textit{Exaptation in science and technology}}
A growing number of scholars in the area of innovation theory propose exaptation as the ultimate source of novelty. They argue that exaptation can explain the emergence of markets, technologies, and technical functions~\cite{dew2016exaptation,cattani2005preadaptation,andriani2017measuring,mokyr1991evolutionary}. In their view, exaptation leads to \emph{technological speciation} or the creation of \emph{technological niches}~\cite{andriani2016exaptation}. In technological speciation processes, new technology develops from the effective transfer of existing knowledge to a new situation, where the transferred knowledge is interpreted and exploited in new ways~\cite{cohen1990absorptive}. Famous cases that illustrate this pattern include Corning Inc., a company that used its long-standing experience on glass engineering to deliver ground-breaking fibre-optic work that has transformed the telecommunications landscape~\cite{watson2017convergence}, or the microwave, which was discovered by chance through the repurposing of parts of a radar system~\cite{rosenman1988serendipity}. Often, a distinction is made between radical and incremental innovations. While radical innovations transform the technological landscape, incremental innovations are minor improvements in existing technologies~\cite{dewar1986adoption}. Exaptation has been associated with radical innovations leading to the creation of new technological niches~\cite{andriani2013exaptation}.
Most empirical studies of exaptation in those contexts have been limited to small-scale or narratives of specific technologies.

In the context of science, contributions to the study of invention have used scientific publications, and metadata, such as author affiliation, organisation, location, and citation linkages to assess novel research outputs. Citation network analysis has been used extensively. Uzzi et al.~\cite{uzzi2013atypical} used co-citation linkages from publications in various fields to differentiate between typical and atypical co-citations. Atypical co-citations are papers that are rarely cited together. They found that high-impact publications were usually not those that had the most atypical or novel combination of ideas, nor those that used typical combinations of ideas, but papers that cite a mix of new and conventional ideas. This result implies that, while originality is a crucial feature of high-impact science, the building blocks for new ideas are often embodied in existing knowledge~\cite{uzzi2013atypical}. Further, papers with high novelty as measured by atypical combinations tend to be less cited at the start, but are more likely to be cited after several years after publication~\cite{stephan2017reviewers}. Other studies, which highlight the role of recombining ideas in driving innovation, suggest that older, seminal works are more likely to inspire ground-breaking science~\cite{kuhn1962structure}. 

The combination of different theories, fields, and tools is also central to interdisciplinary research~\cite{wagner2011approaches}. Interdisciplinarity is likely to lead to more `innovative' research~\cite{thurner2019role}, which is associated with higher levels of citation impact~\cite{lariviere2015long}. Furthermore, the citation influence of papers is enhanced by the thematic distance (i.e., cognitively different fields) from the articles they cite \cite{klavans2013towards}. Yet, such outputs often face more resistance than mainstream publications (i.e., publications drawing mainly on the knowledge of a single field), thus requiring more time to get recognised by the wider scientific community \cite{thurner2019role}. This idea relates to the \quotes{first-mover} advantage where mediocre papers will often receive more citations early on, than a later excellent one \cite{newman2009first}. There is, however, conflicting evidence that interdisciplinary research obtains higher citation rates at the level of journals in natural and medical sciences \cite{levitt2008multidisciplinary}, research departments \cite{rinia2002impact}, and in the field of biomedicine \cite{lariviere2010relationship}. 
This evidence shows that the relationship between novel research - as defined by interdisciplinary combinations - and impact depends on the characteristics of the fields and type of analysis involved \cite{lariviere2010relationship}.

\section{Methodology}
\begin{figure}[t]
    \centering
    \includegraphics[width=0.55\textwidth]{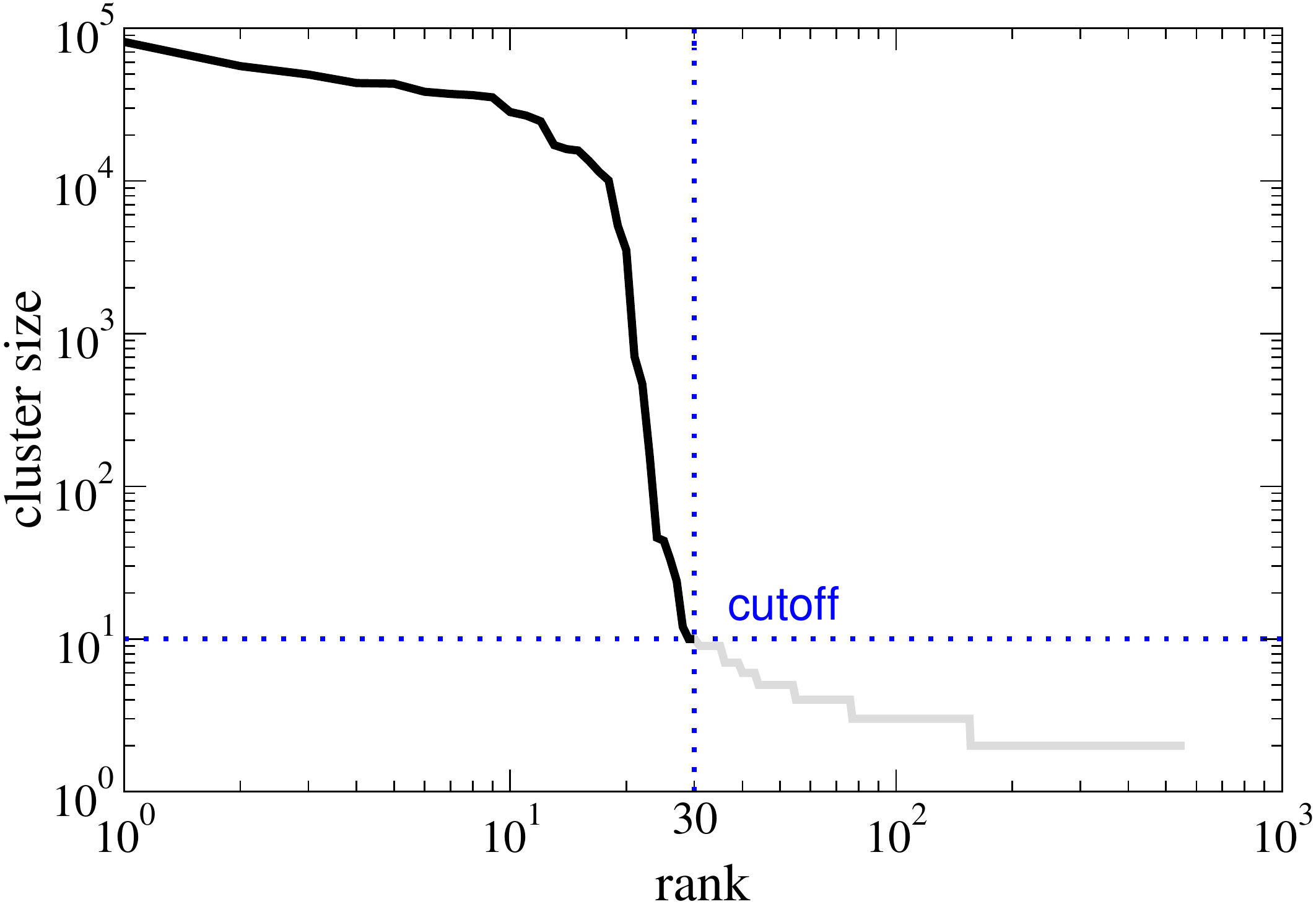}
    \caption{\textbf{Cluster size vs rank.}
    The number of APS articles in a cluster as a function of  the rank of the clusters.
    Clusters were detected by the Louvain algorithm applied to the citation network. 
    We cluster all articles appeared in APS from 1893 to 2017. 
    In our analysis, we remove the articles belonging to clusters with less than 10 articles. 
    These removed clusters constitute a few percent of all articles.}
    \label{fig:cluster_sizes}
\end{figure}
To track scientific progress, we use the APS dataset~\cite{apsdataset}, which includes publications in the leading physics journals since 1893. 
\mm{Following~\cite{blondel2008fast}, we apply the Louvain algorithm to design an alternative classification scheme that clusters the set of all publications (articles and reviews) in the APS between 1893 and 2017 into research areas. The method is based on first determining pairs of publications that cite one another, and second, on clustering publications into a research area. The procedure uses a directed citation network where nodes consist of publications and links consist of citations between publications. Each publication belongs to a unique research area. We disregard the direction of the links in the network and exclude publications without citations.} 

Fig.~\ref{fig:cluster_sizes} shows the distribution of cluster sizes. For several clusters the number of publications is very small. For practical reasons we excluded clusters that have less than 10 publications. This method allows us to examine the influence of publications that belong to a specific field on other papers pertaining to different fields. This information is essential to determine whether a paper has been co-opted by papers in another field. This also allows us to trace the bibliographic properties of the citing publications.
\begin{figure}[t]
  \centering
  \includegraphics[width=0.45\textwidth]{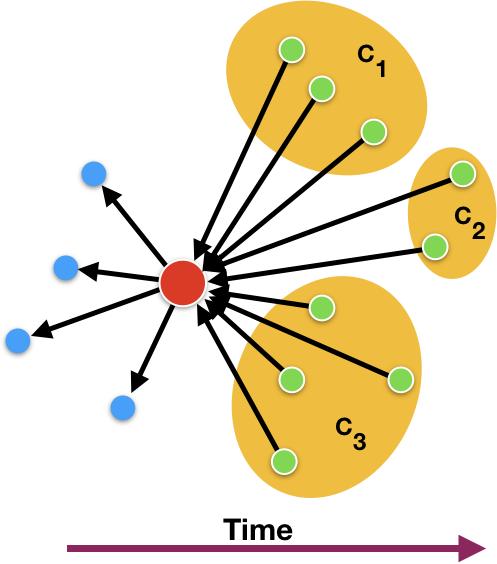}
  \caption{\textbf{Definition of normalised forward entropy cartoon.}
    Circles represent published articles and arrows stand for citations. 
    Only those articles and links are displayed, which are relevant for the entropy definition. The article $a_i$ denoted with the large red circle cites four papers (blue circles at its left side) and has been cited by nine papers (green circles on its right side). These nine articles belong to three different clusters $C_n$ depicted as orange ovals. Denoting with $p_n$ the fraction of articles of cluster $n$ in the whole citation pool of the red article, we have $p_1=1/3$, $p_2=2/9$ and $p_3=4/9$. According to Eq.~(\ref{eq:fwd_entropy}), the normalised forward entropy is then $S_i=(-1/3 \log 1/3 - 2/9 \log 2/9 - 4/9 \log 4/9)/\log 3 \approx 0.966$. The IPR for $a_i$ is $I_i \simeq 2.79$.}
  \label{fig:fwd_entropy}
\end{figure}

\subsubsection*{\textit{Quantifying exaptation}}
To quantify the idea of functionality, we define a quantity that we call {\em forward normalised entropy} (FNE), $S_i$, of a generic article  $a_i$. In Fig.~\ref{fig:fwd_entropy}, we consider the articles citing  $a_i$ (red circle), each belonging to a cluster, $C_n$, here we use all papers published in APS until 2017. Denoting by $A_i$ the set of articles citing $a_i$ and by $C_n$ the clusters in the system, we define $p_{i,n} = |C_n\cap A_i| / |A_i|$. In other words, $p_{i,n}$ is the fraction of articles citing the reference article $a_i$, that belong to cluster $C_n$. We define the normalised forward entropy, $S_i$,  as:
\begin{equation}
  S_i = -\frac{1}{\log N_i}\sum_{n=1}^{N_i} p_{i,n} \log p_{i,n}\ ,
  \label{eq:fwd_entropy}
\end{equation}
\noindent where $N_i$ is the number of clusters for which $p_{i,n}>0$.
We call $S_i$ a forward entropy because it is computed based on articles published after $a_i$ and citing $a_i$.

It gives information on how heterogeneous the composition of the citing articles is in terms of cluster composition. If an article is cited by articles belonging to only one cluster, i.e., it belongs to a well-defined scientific field, $S_i=0$. If a paper has $S_i=1$, then its citations are uniformly distributed among different areas. Therefore, the forward entropy can be thought of as a score for interdisciplinary impact.

To estimate the effective number of clusters from which the paper $a_i$ received citations, we consider the Inverse Participation Ratio (IPR) $I_i$ of article $a_i$: 
\begin{equation}
    I_i = \left(\sum_{n, p_{i,n}>0} (p_{i,n})^{-2}\right)^{-1}.
    \label{eq:ipr}
\end{equation}
Its value approximates the number of effective clusters citing $a_i$. We calculate the forward entropy and IPR for every year, by considering only those articles published in a given year, and citing $a_i$. In this way, we can follow the trajectory of an article over time, keeping track of the scientific areas it belonged to.

\section{Results}
We considered all publications in the APS database until 2017 and selected the top 200 most cited ones. 
The reasons for using highly cited publications are both conceptual and pragmatic. 
First, we assume that exaptation results from the significant adoption or acknowledgement of a paper. 
This implies that a co-opted publication should have a high citation impact. 
Second, a large number of citations enhances the statistical significance of the results. 
To identify distinctive patterns of exaptation, we looked at the yearly number of citations vs.\ the forward normalised entropy (FNE), $S_i$, and the IPR, $I_i$, for all articles. We present a few examples with a well-defined signature of exaptation. 

Before that, let us clarify how this pattern should ideally look like. 
A good candidate article for exaptation, say $a_\mathrm{ex}$, ideally belongs to a well-established field and is disciplinary in nature. 
If the paper initially received citations only by papers in the same scientific sub-field, then its FNE score, $S_i$, would be zero. 
We hypothesise that at some point in time, the number of citations to $a_\mathrm{ex}$ starts increasing 
and possibly remains in the same scientific sub-field. 
At a later stage, the article may start receiving citations by papers from other scientific sub-fields, causing $S_i$ to increase. 
The very fact that papers are co-opted in another scientific sub-domain also brings more citations to $a_\mathrm{ex}$. 
If the new citing domain is highly productive, i.e., with many published papers, then $S_i$ may eventually decline, as most of the citations will now come from the new citing field, overshadowing the original one. 
Eventually, while the citation rate of the paper will start to decrease as a natural consequence of ageing, its FNE, on the contrary, may increase slightly as other fields may become interested in the article and grow in relative importance.

\begin{figure}[t]
    \centering
    \includegraphics[width=0.45\textwidth]{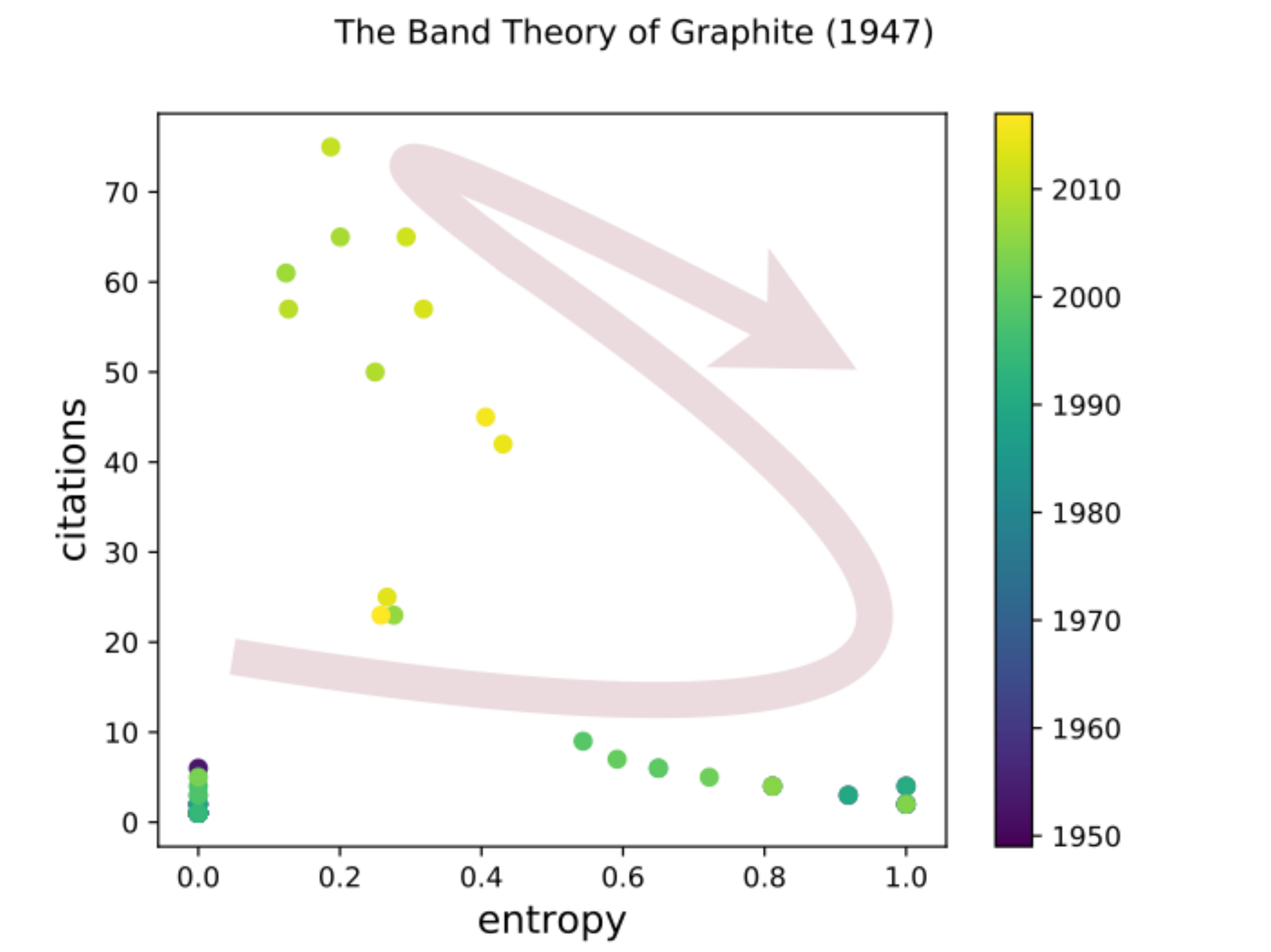}%
    \includegraphics[width=0.45\textwidth]{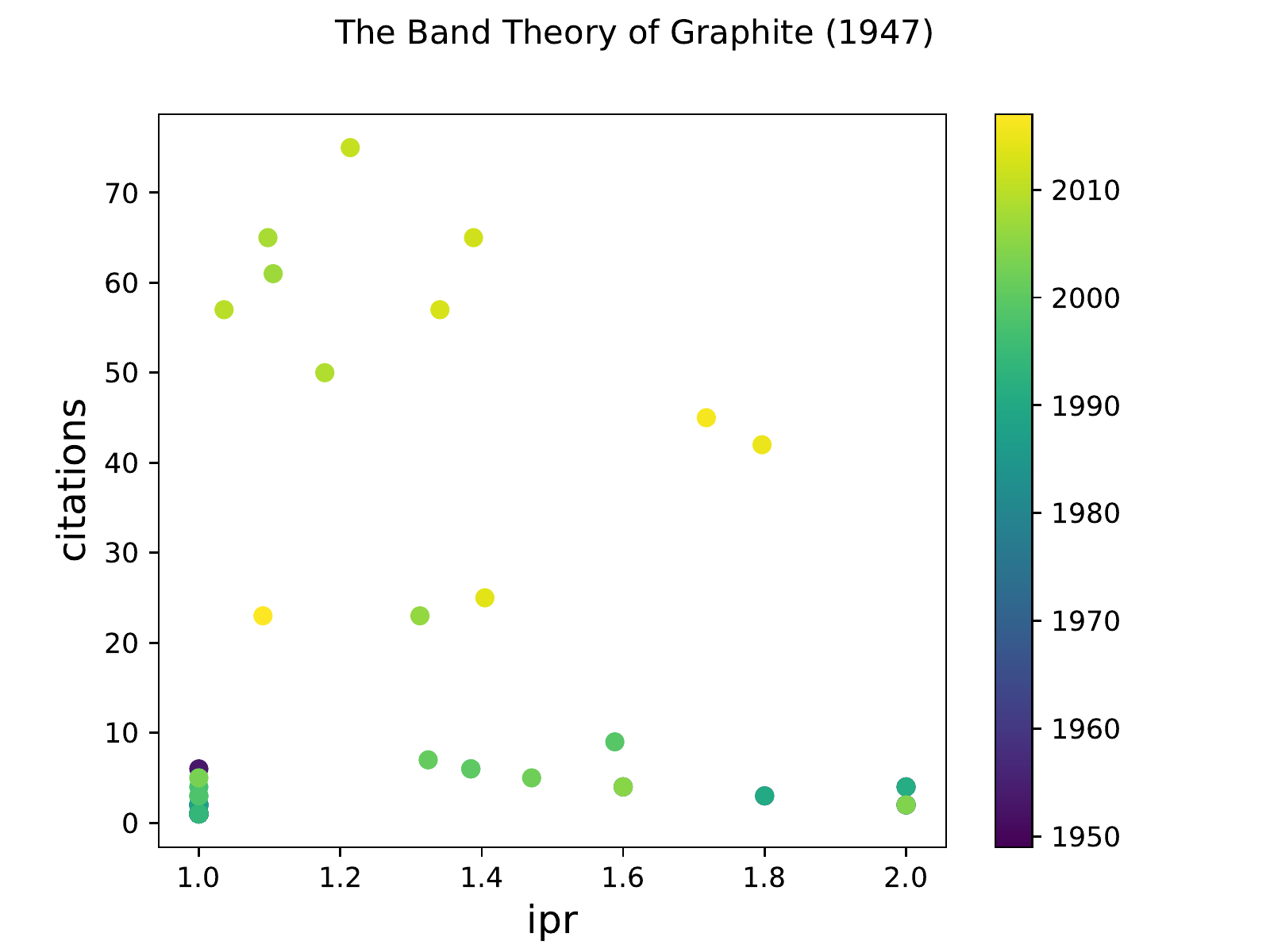}
    \caption{\textbf{A typical pattern for exaptation of an important paper.}
    {Left:} yearly number of citations vs.\ the forward normalised entropy of the paper of Ref.\cite{graphite1947}. Each circle represents a different year, and the time-evolution is marked with a colour code, from dark tones (older) to light ones (newer). The ideal dynamics of citations and entropy over time for an exapted paper is also depicted by a thick arrow.
    The paper starts in one field ($S_i=0$), then is exapted in another field leading to an increase of $S_i$; then, citations grow since the paper gets more popular while $S_i$ decreases, since the new field prevails; eventually, the number of citations decreases as an ageing effect while $S_i$ increases again, since the paper is rediscovered in the older field.
    {Right:} yearly number of citations vs.\ IPR of~\cite{graphite1947}.}
    \label{fig:graphene}
\end{figure}

The paper \emph{The band theory of graphite}, published in 1947~\cite{graphite1947} seems to follow the hypothesised pattern of exaptation. 
Before taking a closer look, let us first dig into the graphene background.
Graphite is a material made up of carbon atoms arranged in parallel hexagonal layers. 
Like the diamond, it is an allotropic form of carbon. 
While graphite is a conductor at room temperature, the diamond is an insulator. 
To understand why the carbon atoms with different crystal geometries are conducting or insulating it is necessary to understand how electrons behave once an electrical potential is applied.
One of the great success of quantum mechanics was the possibility to understand these phenomena.
The electronic structure of simple materials became computable right after the formalism of quantum mechanics was established. 
The geometry of graphite suggests that its electronic properties can be incrementally determined by first analysing a single layer of carbon - what is today known as \textit{graphene} - and then by considering the mutual interaction between layers. As Wallace wrote in his manuscript~\cite{graphite1947}:
\begin{quote}
 Since the spacing of the lattice planes of graphite is large (3.37\AA) compared with the hexagonal spacing in the layer (1.42\AA), a first approximation in the treatment of graphite may be obtained by neglecting the interactions between planes, and supposing that conduction takes place only in layers.
\end{quote}
\noindent 
Wallace did not consider the possibility  that a hexagonal layer of carbon atoms could exist by itself and used it as a mere tool to solve the ``more complicated'' problem for the three-dimensional graphite. 
As recently pointed out in a review essay on graphene~\cite{graphene_review2009}:
\begin{quote}
 It was P.~R.~Wallace, who in 1946 wrote the first papers on the band structure of graphene and showed the unusual semi-metallic behaviour in this material (Wallace, 1947). At that point in time, the thought of a purely 2D structure was a mere fantasy and Wallace's studies of graphene served him as a starting point to study graphite, a very important material for nuclear reactors in the post-World War II era.
\end{quote}
\begin{figure}[t]
    \centering
    \begin{tabular}{cc}
    Word cloud for the graphene cluster&
    Word cloud for the ESP cluster\\
    \includegraphics[width=0.48\textwidth]{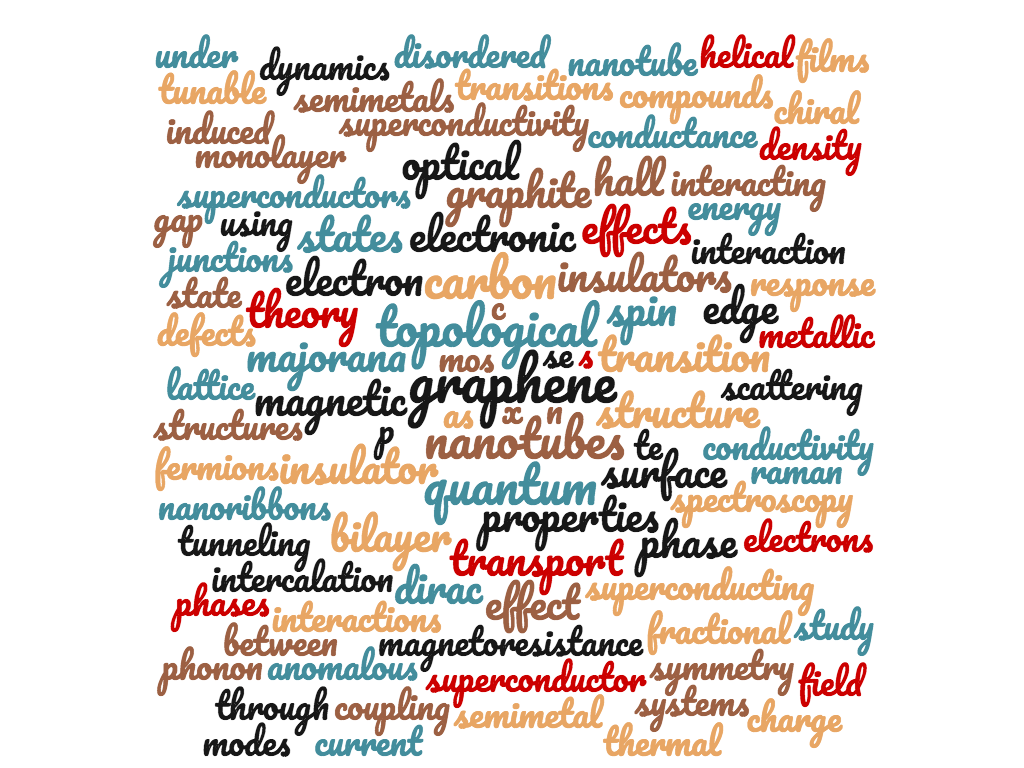}&
    \includegraphics[width=0.48\textwidth]{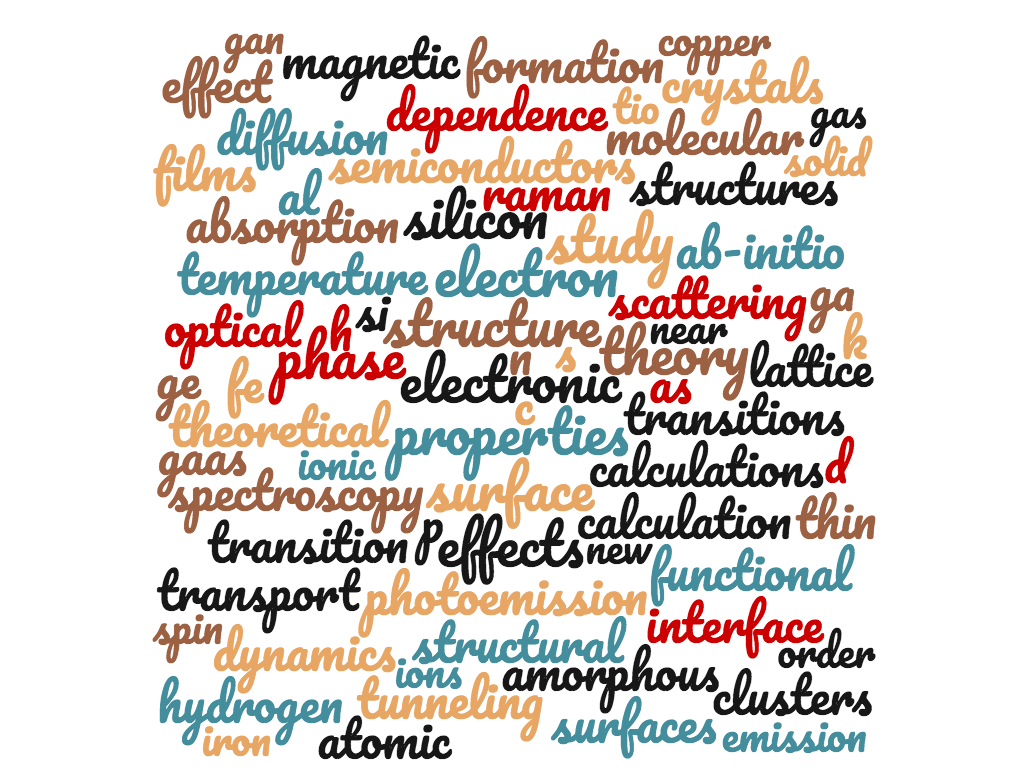}
    \end{tabular}
    \caption{\textbf{Word clouds of the two communities citing Wallace's paper.}
    Word clouds~\cite{wordcloud} were generated by merging all the titles belonging to the main two clusters citing the graphite/graphene article. Stop words, numbers and punctuation marks were removed. Font size is proportional to the logarithm of the number of occurrences of words. The composition of the main cluster (graphene) with 613 articles citing Wallace's graphite article \protect\cite{graphite1947} is shown on the left; the second important cluster (electronic structure properties) with 33 articles is on the right.
    }
    \label{fig:wordclouds}
\end{figure}
Let us see now why the seminal paper of Wallace on graphene can be considered an example of exaptation according to the pattern highlighted above. Figure~\ref{fig:graphene}, shows two panels of the yearly number of citations vs.\ the forward normalised entropy (FNE) $S_i$ (left), and the IPR, $I_i$ (right). 
The paper starts with a small number of citations and a very small FNE (bottom-left corner of the plot in the left panel). 
After that, the pattern follows a very peculiar ``S-shaped'' trajectory, mirroring the following three situations: (i) the adoption by another field leading to an increase of $S_i$, (ii) the growth in the number of citations while $S_i$ decreases, (iii) the decrease of the number of citations while $S_i$ increases again.

This suggests that Wallace's article that was meant to belong to the field of electronic structure properties (ESP), was later co-opted in the graphene research field. 
According to the cluster analysis, the ESP field contains approximately 12,000 articles and the much more recent graphene area contains about 5,000 articles. 
Despite the larger number of articles in ESP, Wallace's paper has been cited mostly by the graphene community (613 times), and much less from the  ESP (33 times). 
The forward normalised entropy of this paper follows the exaptation pattern we assumed. 
Its IPR pattern (Fig.~\ref{fig:graphene} (right panel)) demonstrates how it switched from the ESP to the graphene community. 
To visualise the differences between the two ESP and graphene communities, in Fig.~\ref{fig:wordclouds} we show two word-clouds from the articles citing Wallace's paper: from the graphene community  (left) and the ESP one (right). 

\begin{figure}[t]
    \centering
    \includegraphics[width=0.95\textwidth]{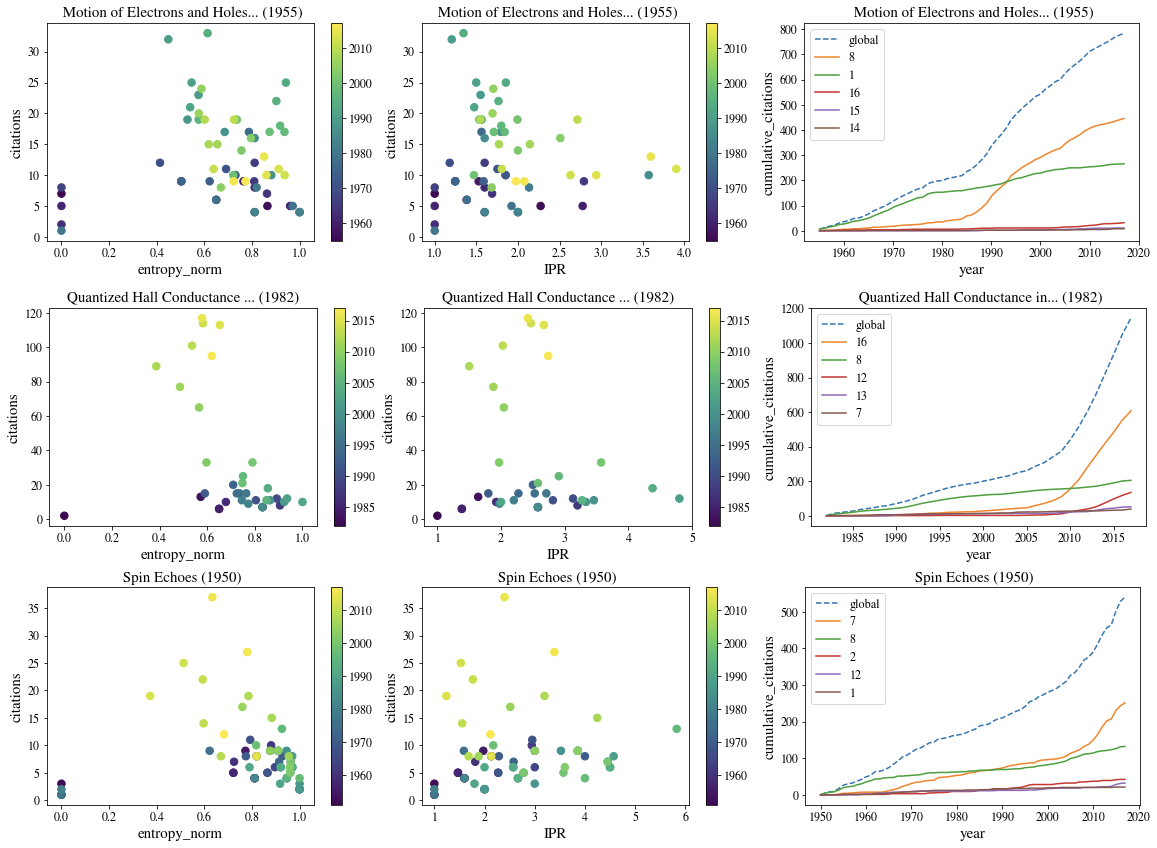}
    \caption{\textbf{More examples of exaptation patterns.} Three articles are shown: Refs.~\protect\cite{electrons_and_holes,quantized_hall,spin_echoes}, one in each row. In the left and centre panels, we display their citation versus FNE and IPR, respectively. Patterns are similar to what was found in Fig.~\ref{fig:graphene}. In panels to the right, we show the time evolution of their cumulative citations in the respective clusters, whose id number is given in the legend. Their vertical order reflects their ranking in 2017, i.e., their relative importance in citing the selected paper.}
    \label{fig:examples}
\end{figure}

We found 10 additional articles whose FNE and IPR patterns are similar to those of the graphite/graphene paper. Among those, we mention two notable examples, i.e., \emph{Motion of Electrons and Holes in Perturbed Periodic Fields}~\cite{electrons_and_holes} and \emph{Quantized Hall Conductance in a Two-Dimensional Periodic Potential}~\cite{quantized_hall}. 
A third article, that was not in the list of the most 200 cited papers but was highly cited from outside the APS, was chosen on the basis of our personal experience, \emph{Spin Echoes}~\cite{spin_echoes}. 
The first ~\cite{electrons_and_holes}, is mainly cited by articles belonging to two clusters,  the ``Quantum dots / Quantum wells'' cluster (QDQW) and the ESP cluster. These clusters are denoted by id=8 and id=1 respectively, and their mutual importance is sketched in the plots on the right panels of Fig.~\ref{fig:examples}.  
Note how after 1980, the QDQW cluster starts to become more important than ESP, so that~\cite{electrons_and_holes} acquires more importance in that field. 
A similar situation occurs with reference ~\cite{quantized_hall} (second row of the panel), where now the graphene cluster (id=16) competes with the QDQW cluster. Interestingly, from year 2010 on, also the ``Bose-Einstein condensate'' cluster (id=12) starts to get some importance.

The third paper on spin echoes was chosen since we expected its importance on Nuclear Magnetic Resonance (NMR) would become prominent in time (third row of the panel). 
We do not observe a clear NMR cluster in APS, rather we find an exaptation pattern, where the ``Entanglement'' cluster (id=7) emerges on the detriment of the QDQW cluster (id=8), as shown in the bottom right panel. 
We think that the NMR community cites this paper from outside the APS community, i.e., by papers not published by the APS journals.

\begin{figure}[t]
  \centering
  \includegraphics[width=0.42\textwidth]{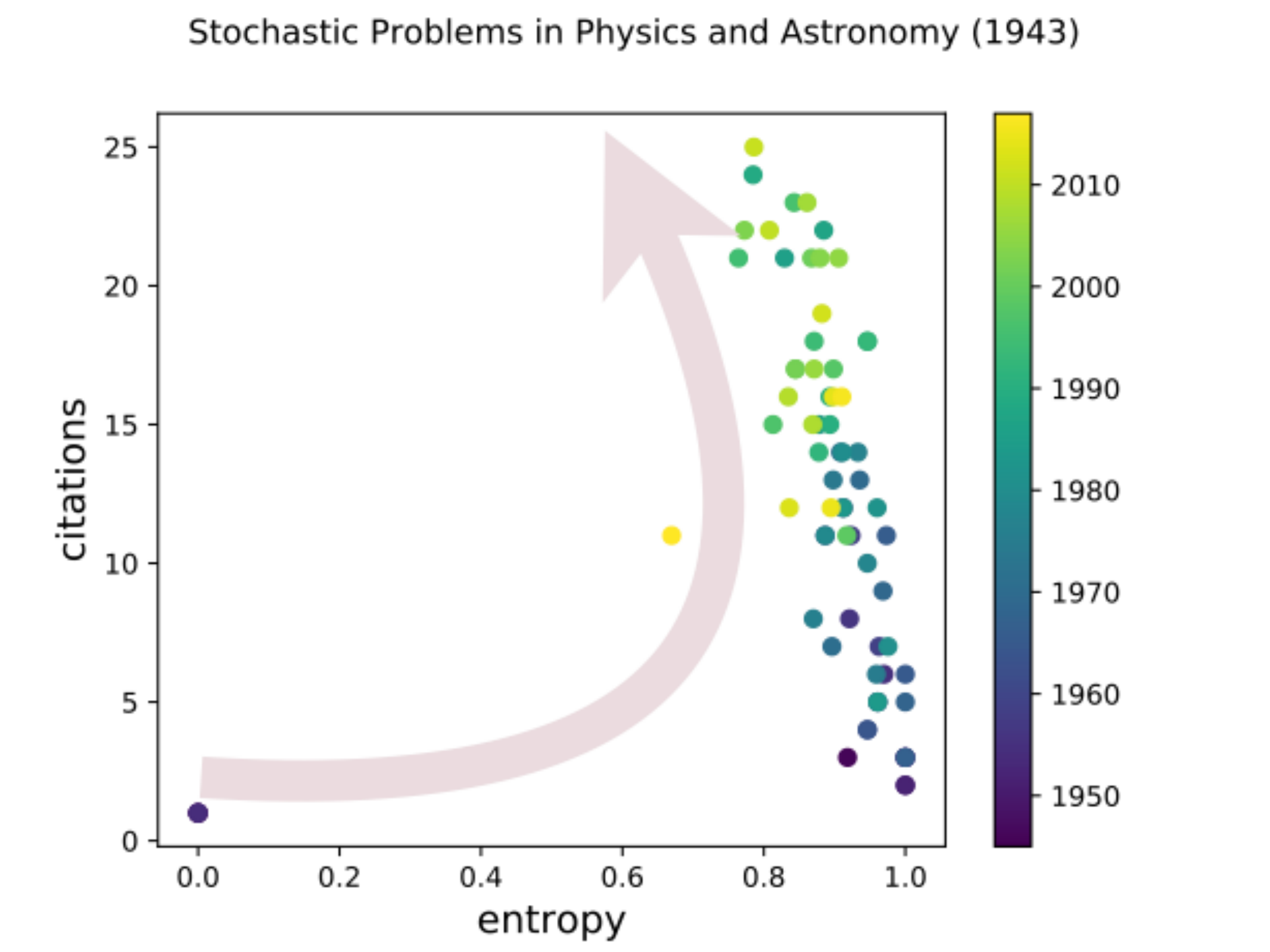}%
  \includegraphics[width=0.42\textwidth]{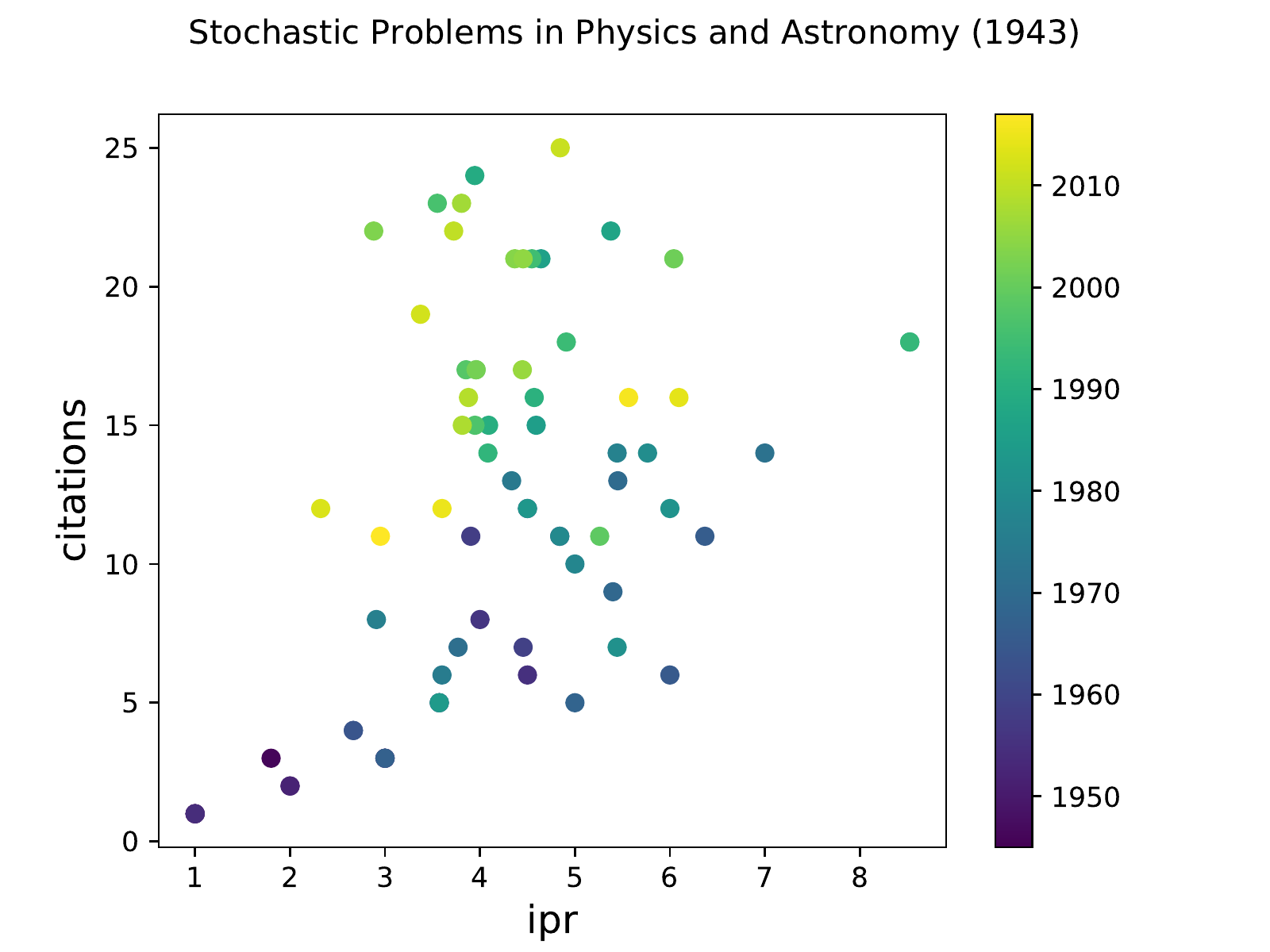}
\caption{\textbf{FNE and IPR pattern for a typical review article in APS.} Review articles exhibit a distinctive dynamical behaviour slightly different than the exaptation patterns discussed in Figs. 3. and 5. Left: the forward normalised entropy increases quickly and remains at high values over time, while the number of citations per year grows. Right:  the number of clusters citing a review article grows over time showing an increased interest in several sub-disciplines.}
  \label{fig:review}
\end{figure}

Although not directly related to the problem of detecting exaptation in APS articles, it is worthwhile looking at the time evolution of the FNE and the IPR for review articles. We observe the same kind of pattern consistently in all highly cited APS review papers. As an example, we show the Chandrasekhar's review \emph{Stochastic problems in physics and astronomy}~\cite{review1943}; the corresponding plots for FNE and IPR are reported in Fig.~\ref{fig:review}.
Both FNE and IPR increase over time, witnessing the number of different scientific fields interested in the review. The manuscript itself~\cite{review1943} features four papers, the problem of random flights, the theory of the Brownian motion, probability after effects, and probability methods in stellar dynamics. The IPR  value reaches the value four soon and oscillates around it in time. On the other hand, the FNE, after reaching its maximum value, starts to decrease, mainly because one citing field increases its importance over the others. The IPR and FNE behaviour of highly cited reviews is essentially different from the patterns found for co-opted articles presented in Figs.~\ref{fig:examples} and \ref{fig:graphene}. This finding further reassures us of the robustness of the proposed indicators, as well as of the overall methodology. 
\section{Discussion}

The concept of exaptation has not been appropriately quantified. 
Only very few systematic quantitative analyses are available in the literature. 
Here we focused on the exaptation phenomena in the framework of scientific progress as it can be observed in the APS article collection database. 
Our approach consists of looking for signatures of exaptation in the adoption (as reflected in citation patterns) of specific results by other scientific communities than the original scientific area of belonging. 
To make the analysis quantitative, we propose a method based on two main components: clustering of publications based on citation relations, and two observables, the Forward Normalised Entropy (FNE) and the Inverse Participation Ratio (IPR). 
These measures allow to single out patterns that are related to the exaptation phenomena in scientific evolution. 

\mm{The evolution of feathers is a classical example of exaptation. 
Feathers were initially adapted to protect against thermal excursions and were later co-opted for flying. 
As soon as birds learn to fly, their fitness improves since they have higher chances of surviving predators. 
We extend this reasoning with a thought experiment ({\em Gedankenexperiment}) to the exaptation of scientific knowledge, where citations represent the fitness and the inverse participation ratio represents the number of  functions a publication acquires over time.
The  example of graphene constitutes a popular instance of exaptation in physics, particularly in contributing to the development of two related sub-fields, electronic structure properties, and graphene. 
It shows that the `survival' of a field depends on how pre-existing concepts are re-used or applied by communities in new domains resulting in new functionalities. 
}

\mm{These results should be seen as preliminary steps towards a quantitative theory of exaptation; they show that is in principle possible to quantify exaptation, once it becomes possible to get a quantitative handle on the evolving context of a field. 
A better theory of exaptation is necessary to explain how something emerges and evolves by exploring its \emph{creative potential}~\cite{andriani2016exaptation}. 
The creative potential has been described as the functional shift of a particular component that could open up an evolutionary path different from its original and perhaps intentional trajectory~\cite{andriani2016exaptation,kauffman1993origins}. 
We think that this paper could stimulate a new wave of studies in this promising area of scientific research. }


\section*{Acknowledgements}
This work was supported by the Austrian Research Promotion Agency FFG under grant 857136.

\bibliographystyle{plain}
\bibliography{exaptation}
\end{document}